\documentclass[footinbib,aps,prb,reprint,preprintnumbers,superscriptaddress,amsmath,amssymb,bibnotes,longbibliography]{revtex4-2}
\pdfoutput=1
\usepackage{graphicx}
\usepackage[colorlinks,linkcolor=blue,anchorcolor=blue,citecolor=blue,urlcolor=blue,filecolor=blue,menucolor=blue,runcolor=blue]{hyperref}
\usepackage{gensymb}
\usepackage{epstopdf}

\def\cred{%
}

\newcommand{\av}{\vec{a}}
\newcommand{\bv}{\vec{b}}
\newcommand{\cv}{\vec{c}}

\graphicspath{{figures/}{../figures/}}

\begin{document}

\title{Geometrical frustration and incommensurate magnetic order \\ in Na$_3$RuO$_4$ with two triangular motifs}

\author{Vera P. Bader}
\email{vera.bader@uni-a.de}
\affiliation  {Experimental Physics VI, Center for Electronic Correlations and Magnetism, University of Augsburg, 86159 Augsburg, Germany}

\author{Clemens Ritter}
\affiliation  {Institut Laue-Langevin, 38042 Grenoble Cedex 9, France}

\author{Elias Papke}
\affiliation  {Experimental Physics VI, Center for Electronic Correlations and Magnetism, University of Augsburg, 86159 Augsburg, Germany}

\author{Philipp Gegenwart}
\affiliation{Experimental Physics VI, Center for Electronic Correlations and Magnetism, University of Augsburg, 86159 Augsburg, Germany}

\author{Alexander A. Tsirlin}
\email{altsirlin@gmail.com}
\affiliation{Experimental Physics VI, Center for Electronic Correlations and Magnetism, University of Augsburg, 86159 Augsburg, Germany}
\affiliation{Felix Bloch Institute for Solid-State Physics, University of Leipzig, 04103 Leipzig, Germany}

\date{\today}

\begin{abstract}

Incommensurate magnetic order in the spin-3/2 antiferromagnet Na$_3$RuO$_4$ is uncovered by neutron diffraction combined with \textit{ab initio} calculations. The crystal structure of Na$_3$RuO$_4$ contains two triangular motifs on different length scales. The magnetic Ru$^{5+}$ ions form a lozenge (diamond) configuration, with tetramers composed of two isosceles triangles. These tetramers are further arranged in layers, such that an effective triangular lattice is formed.
The tetramers are nearly antiferromagnetic but frustration between them leads to an incommensurately modulated magnetic structure described by the propagation vector $\vec{k}=(0.242(1), 0, 0.313(1))$. We show that the long-range Ru--O--O--Ru couplings between the tetramers play a major role in Na$_3$RuO$_4$ and suggest an effective description in terms of the spatially anisotropic triangular lattice if the tetramers are treated as single sites.
\end{abstract}

\maketitle


\section{Introduction}
\begin{figure*}
\centering
\includegraphics[angle=0,width=0.8\textwidth]{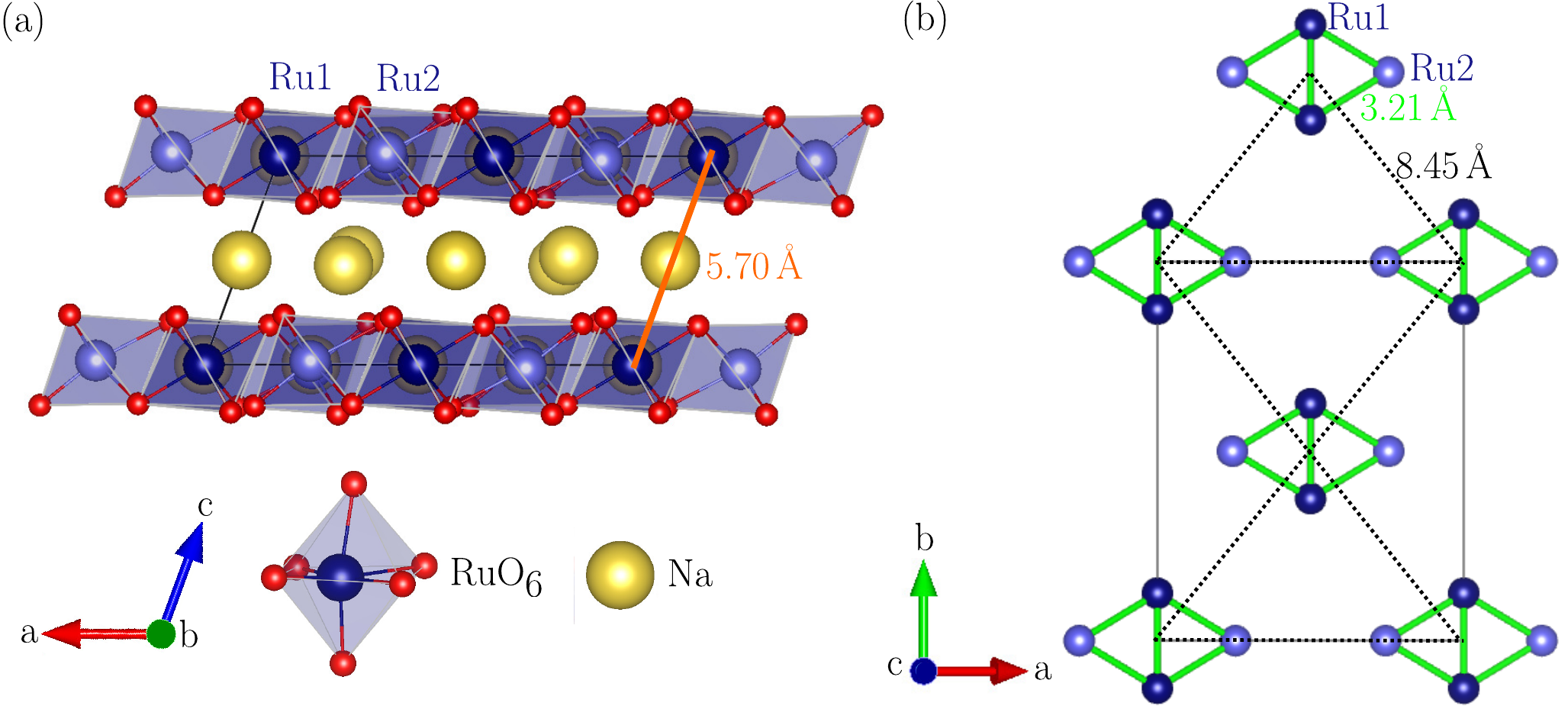}
\caption{\label{structure} Crystal structure of Na$_3$RuO$_4$ at 300\,K. (a) View along the $b$ axis. The magnetic layers containing ruthenium are separated by non-magnetic sodium layers. (b) View along the $c$ axis. The RuO$_6$ octahedra form tetramers, which are composed of two isosceles triangles (drawn in green). The tetramers themselves are arranged on a triangular lattice (shown by the black dotted lines). Na and O were omitted for clarity. The structures were drawn with \texttt{VESTA}.}
\end{figure*}

Ru$^{4+}$ compounds show a plethora of interesting properties like superconductivity in Sr$_2$RuO$_4$ \cite{Maeno1994}, ferromagnetism in SrRuO$_3$ \cite{Kanbayasi1976}, Pauli paramagnetism in BaRuO$_3$ \cite{Jin2008}, and the $J=0$ electronic state that may serve as a platform for excitonic magnetism \cite{Chaloupka2019}.
Other oxidation states of Ru can be peculiar too. Strong electronic correlations render Ru$^{4+}$-based Pr$_2$Ru$_2$O$_7$ an antiferromagnetic Mott insulator, whereas Ru$^{5+}$-based Ca$_2$Ru$_2$O$_7$ is a paramagnetic metal \cite{Kaneko2021}. Ruthenium compounds are also instrumental in studying quantum and frustrated magnets with different values of the local spin \cite{marjerrison2016,prasad2020,schnelle2021,hiley2014,suzuki2019,kumar2019,ortiz2023}.\\
\indent In the present case, we concentrate on a Ru$^{5+}$ (4$d^3$) system in which the magnetic ions carry a spin of $S=3/2$ with the quenched orbital moment. The layered Ru$^{5+}$ compound Na$_3$RuO$_4$ hosts a number of different actors, including frustration, low-dimensionality, and \mbox{Ru-O} covalency. Na$_3$RuO$_4$ first synthesized by Darriet \textit{et al.}~\cite{Darriet1974} crystallizes in the space group $C2/m$ (no. 12), which was later confirmed through powder neutron diffraction \cite{Regan2005}. The Ru$^{5+}$O$_6$ octahedra form tetramers that are composed of two isosceles Ru triangles. {\cred The tetramers themselves build layers, which are stacked along the crystallographic $c$ axis (Fig.~\ref{structure}). Geometrically, these tetramers form a slightly deformed square lattice with the nearest-neighbor distance of 8.45\,\r A (half of the $ab$-face diagonal). The distance of 11.03\,\r A between the tetramers along the $a$-direction is slightly longer, but it corresponds to the similar O--O contacts of about 3.5\,\r A for the adjacent tetramers. One can then envisage that the tetramers build} a triangular lattice (Fig.~\ref{structure}b), a description that we put forward in our present study. In Na$_3$RuO$_4$, the triangular geometry, the simplest motif for geometrical frustration, is thus found on two different length scales of the crystal lattice.\\
\indent The tetramers in the planes and the tetramers in adjacent layers are separated by Na. At first glance, the large spatial separation between the tetramers suggests that they should be almost decoupled. This scenario was put forward by Drillon \textit{et al.}~\cite{Drillon1977} who was able to describe magnetic susceptibility data using only two isotropic intratetramer exchange couplings. At low temperatures, the individual tetramers should approach their singlet ground state without long-range magnetic order. Later, however, M\"ossbauer spectroscopy \cite{Gibb1980} and neutron diffraction \cite{Regan2005, Haraldsen2009} detected the formation of magnetic order in Na$_3$RuO$_4$ below $28-30$\,K with the concomitant magnetic excitations that were observed in inelastic neutron scattering and attributed to acoustic and optical spin waves. Their damping with temperature indicates an intertetramer coupling~\cite{Haraldsen2009}. Additionally, heat capacity data revealed two successive phase transitions at 25 and 28\,K, respectively~\cite{Haraldsen2009}. These transitions either do not appear or could not be resolved in the previous magnetization data that showed one transition only~\cite{Regan2005,Haraldsen2009}.\\
\indent The nature of the ordered state in Na$_3$RuO$_4$ remains unknown. 
It was shown over time that the tetramers are not decoupled, but how exactly does the magnetic structure look like, and what is the nature of intertetramer couplings? In this work, we seek to address these questions. Using powder neutron diffraction (NPD) and \textit{ab initio} calculations, we determine the magnetic structure of Na$_3$RuO$_4$ and underlying exchange couplings. We uncover the incommensurate nature of the magnetic order and show how this incommensurability arises from the two length scales of the magnetic frustration. Triangular structural motifs play crucial role in stabilizing the incommensurate order and prevail over the intuitively anticipated tetramer physics.

\section{Methods}
Powder samples of Na$_3$RuO$_4$ were synthesized similar to the procedure described in Ref.~\cite{Regan2005}. NaOH and pre-dried RuO$_2$ (2\,h at 700\,\degree C) were mixed in a molar ratio 3.1:1 and ground in an Ar~glovebox. Here, a small excess of NaOH was used to avoid RuO$_2$ impurities, which are difficult to remove afterwards. The resulting powder was filled in an alumina combustion boat and placed in a horizontal tube furnace. The sample was heated to 500\,\degree C in 2\,h in an O$_2$~flow (20\,sccm) and kept at this temperature for 18\,h. After an intermediate grinding step, the sample was again loaded into the tube furnace. The second annealing was done in the mixed Ar and O$_2$ flow (3:1, 20\,sccm) with the heating to 650\,\degree C in 2\,h and the subsequent dwelling time of 18\,h. The resulting black powder was again ground in the glovebox. 

All experiments were done on powder samples. Sample quality was checked by powder x\mbox{-}ray diffraction (PXRD) performed in Bragg-Brentano geometry with CuK$_{\alpha}$ radiation using the diffractometer \texttt{EMPYREAN} from Panalytical. No impurity phases were detected. For heat capacity measurements a pellet was pressed and tempered. The tempering was conducted in the same atmosphere as used in the second annealing step. The sample was heated to 650\,\degree C in two days, kept at this temperature for three days, and slowly cooled to room temperature in three days. For neutron diffraction measurements the tempered powder from seven different batches with the total mass of 3.45\,g was ground together in an Ar~glovebox. Because the compound is hygroscopic, the sample was loaded into a vanadium container and sealed inside the glovebox.\\
\indent
For VSM magnetization measurements in the \texttt{MPMS3} from Quantum Design, a 7.34\,mg powder sample was filled into a plastic capsule mounted on a brass sample holder. The magnetization was measured at 1, 3, and 7\,T in the temperature window between 2 and 300\,K. The field-dependent magnetization was measured in the field window between 0 and 7\,T for different temperatures.\\

\begin{figure*}
\includegraphics[angle=0,width=\textwidth]{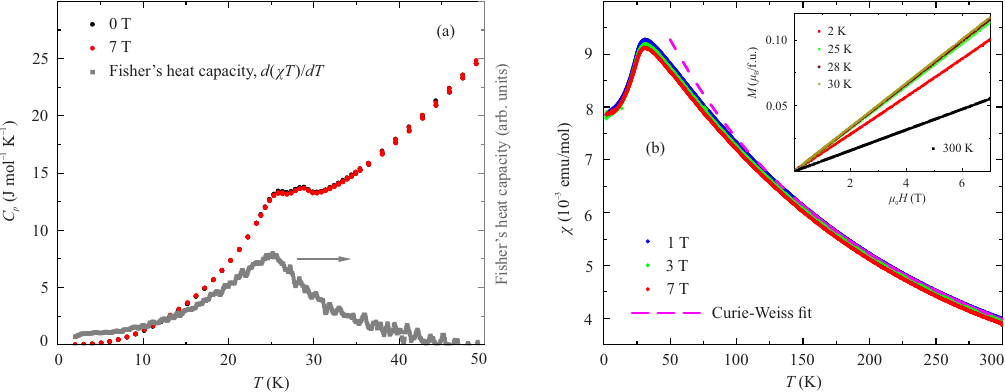}
\caption{\label{HCMagn} Thermodynamic properties of Na$_3$RuO$_4$. (a) Heat capacity data in zero field and at 7\,T. Two successive phase transitions at 26\,K and 29\,K are visible. The lower-temperature transition is also visible in Fisher's heat capacity, $d(\chi T)/dT$, indicated in gray. (b) Temperature-dependent magnetic susceptibility data for three different fields. An antiferromagnetic phase transition is observed at around 30\,K. The high-temperature data are described accurately by the Curie-Weiss fit shown in magenta. The inset shows linear field dependence of the magnetization at several temperatures.}
\end{figure*}

\indent
The heat capacity was measured in the \texttt{PPMS} from Quantum Design with the relaxation method. Before the measurement, the addenda of the platform and low-temperature grease was measured in zero magnetic field and at 7\,T. A 5.41\,mg piece of the tempered pellet was then measured in 0 and 7\,T between 2 and 50\,K.\\
\indent NPD data were collected on the high-resolution diffractometer D2B and on the high-intensity diffractometer D20 of the ILL (Grenoble, France) with the wavelengths 1.594\,\r A and 2.415\,\r A, respectively. The D2B measurements were performed at several temperatures between 1.5 and 300\,K to detect possible structural changes, whereas the D20 measurements were performed in order to reveal a larger number of magnetic reflections. Crystal and magnetic structures of Na$_3$RuO$_4$ were refined in \texttt{Jana2006}~\cite{jana2006} and \texttt{FullProf}~\cite{FullProf1993}, respectively. VESTA~\cite{Vesta2011} and FPStudio were used for visualization

\textit{Ab initio} band-structure calculations were performed in the framework of density functional theory (DFT) with the Perdew-Burke-Ernzerhof type of the exchange-correlation potential \cite{pbe96}. The magnetic structure of Na$_3$RuO$_4$ was modeled using an effective spin Hamiltonian,
\begin{equation}
 \mathcal H=\sum_{\langle ij\rangle}J_{ij}\mathbf S_i\mathbf S_j
\end{equation}
where the summation is over bonds $\langle ij\rangle$ and $S=\frac32$. Exchange couplings $J_{ij}$ were evaluated by a mapping procedure~\cite{xiang2011,tsirlin2014} {\cred using the crystallographic unit cell of Na$_3$RuO$_4$ and the enlarged primitive cells, $(\av+\bv)\times(\av-\bv)\times\cv$ as well as $(\av+\bv)/2\times(\av-\bv)/2\times 2\cv$}~\cite{suppl}. Total energies for different spin configurations were obtained on the relativistic DFT+$U$+SO level in \texttt{VASP}~\cite{vasp1,vasp2} with the mean-field correction for the correlation effects in the Ru $4d$ shell using the on-site Coulomb repulsion $U_d=3$\,eV and Hund's coupling $J_d=0.5$\,eV~\cite{granas2014,hariki2017}. 

\section{Experimental results}
\subsection{Thermodynamic Measurements}
The temperature-dependent heat capacity between 2\,K and 50\,K in zero magnetic field and at 7\,T is shown in Fig.~\ref{HCMagn}(a). Clearly visible are two adjacent phase transitions at 26\,K and 29\,K. The transitions are field independent up to at least 7\,T. Our findings agree well with the data from Ref.~\cite{Haraldsen2009}. We want to mention that in Ref.~\cite{Wang2014} the same double peak of the heat capacity was observed too, but assigned to Na$_2$RuO$_3$. Here we join the opinion of Veiga~\textit{et al.} that the data were miscategorized \cite{Veiga2020}. Na$_3$RuO$_4$ forms easily as side product if traces of oxygen are present during the synthesis of Na$_2$RuO$_3$.\\

\begin{figure*}
\includegraphics[angle=0,width=\textwidth]{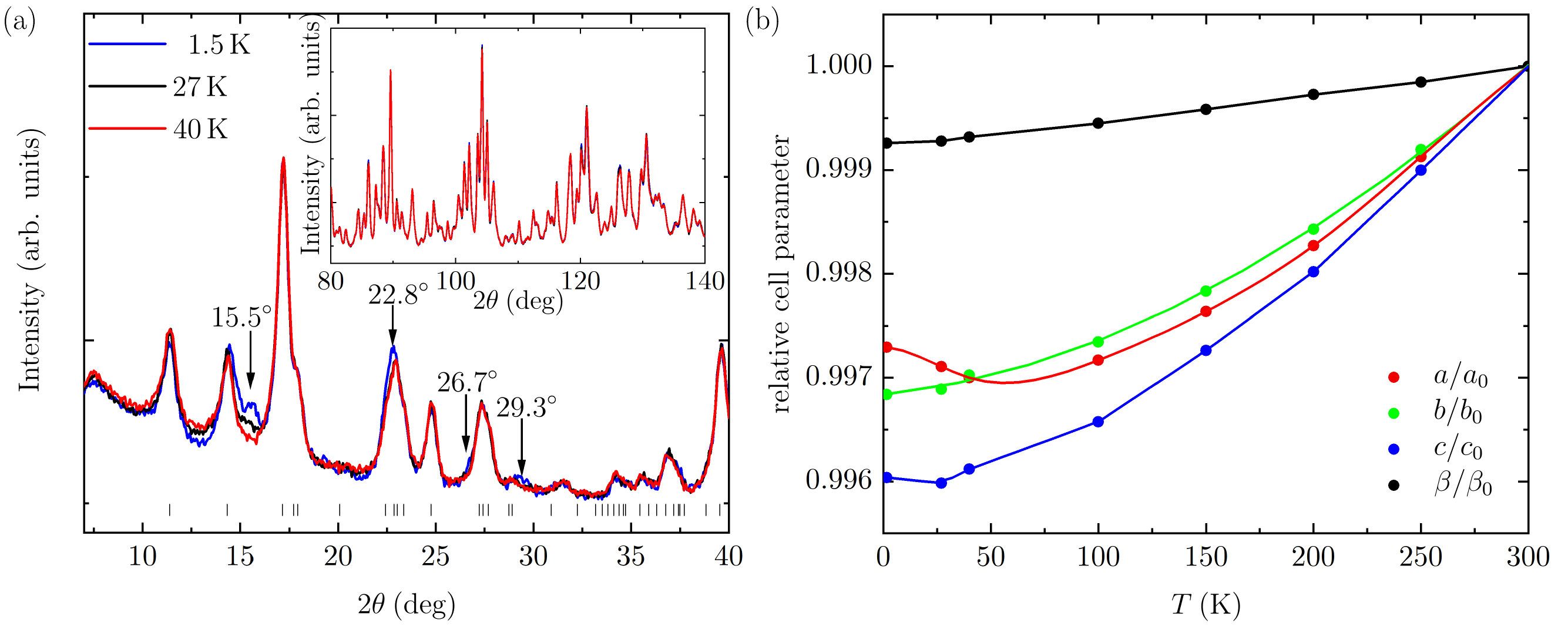}
\caption{\label{D2B} NPD data collected at the high-resolution instrument D2B. (a) NPD pattern at low angles for the temperatures of 1.5, 27, and 40\,K. Magnetic peaks visible at 1.5\,K and partly at 27\,K are marked with arrows. The nuclear peak positions are marked with ticks. The inset shows the data at high angles. No structural phase transition is observed. (b) Temperature-dependent lattice parameters scaled to their values at 300\,K. The error bars are smaller than the symbol size. The lines are guides for the eye.}
\end{figure*}

\indent Both transitions are visible in the magnetic susceptibility as well. While the upper one appears as a maximum in $\chi(T)$ with the N\'{e}el temperature $T_{\text N}\approx 30$\,K [see Fig.~\ref{HCMagn}(b)], the lower one is seen as a maximum in Fisher's heat capacity, $d(\chi T)/dT$ [see Fig.~\ref{HCMagn}(a)], corresponding to the inflection point of the magnetic susceptibility \cite{Fisher1962}. The Curie-Weiss fit, $\chi=\chi_0+(C/(T-\theta))$, in the temperature range of 150 to 300\,K describes the data accurately and returns the Curie-Weiss temperature $\theta$ of $-161.5(5)$\,K for the data collected at 1\,T and an effective moment of $4.07(2)\,\mu_{\text{B}}$, which is close to the spin-only value $3.87\,\mu_{\text{B}}$ expected for a $S=3/2$ ion. The fit parameters for the data at higher fields differ by less than 2\%. To account for the diamagnetic contribution from the sample $\chi_{\text{dia}}\approx -9\cdot 10^{-5}$\,emu/mol (Pascal's constants taken from Ref.~\cite{Bain2008} and assuming similar constants for the different oxidation states of Ru) and the sample holder, a temperature independent constant $\chi_0$ was added to the Curie-Weiss fit \mbox{($\chi_0=-4.97(8)\cdot 10^{-4}$\,emu/mol)}.\\
\indent The ratio \mbox{$\theta/\text{T}_\text{N} \approx 5.4$} indicates a suppression of the magnetic order and may be a fingerprint of the underlying magnetic frustration, although a weak coupling between the tetramers could lead to the reduction of $\text{T}_\text{N}$ as well. We will discuss this further while analyzing the magnetic structure. The magnetization versus field, intertwined with the magnetization versus temperature, is shown in the inset of Fig.~\ref{HCMagn}(b). Due to the antiferromagnetic nature of the transition, the slope of the $M(H)$ curves first increases with increasing temperature from 2 to 30\,K and decreases after the compound enters the paramagnetic regime.

\subsection{Neutron Diffraction}
\subsubsection{Crystal Structure}
The lattice parameters obtained from the Rietveld refinement with the space group $C2/m$ (no. 12) using the NPD data collected at 300\,K are $a=11.0286(1)$\,\AA, $b=12.8148(2)$\,\AA, $c=5.7048(1)$\,\AA, and $\beta=109.905(1)$\degree\ in a perfect agreement with Ref.~\cite{Regan2005}. We also find the same atomic positions as in the previous studies~\cite{suppl}.
\\
\indent The data obtained at the high-resolution instrument D2B at low temperatures are shown in Fig.~\ref{D2B}. Neutron diffraction data were collected above the transition temperatures at 40\,K, in the transition region at 27\,K, and at the base temperature of 1.5\,K. At 1.5\,K, in the magnetically ordered state, no additional peaks and no peak splitting are visible in the data at high angles and hence any changes in the structural symmetry can be ruled out [see the inset in Fig.~\ref{D2B}(a)]. We can thus exclude a structural phase transition within the resolution of our measurement and suggest that both transitions must have predominantly magnetic character.
Indeed, temperature-dependent lattice parameters given in Fig.~\ref{D2B}(b) do not show any significant anomalies but decline monotonically with decreasing temperature. Only the $a$ parameter shows a slight upturn below 50\,K, indicative of the weak magnetoelastic coupling, which is typical for transition-metal compounds~\cite{senn2013,reschke2020}. Already in this setup magnetic Bragg peaks are visible. They are indicated by arrows in Fig.~\ref{D2B}(a). One magnetic peak at 15.5\degree{} develops at 27\,K. At the lowest temperature four peaks are perceptible.

\subsubsection{Magnetic Structure}
\begin{figure}
\includegraphics[angle=0,width=0.48\textwidth]{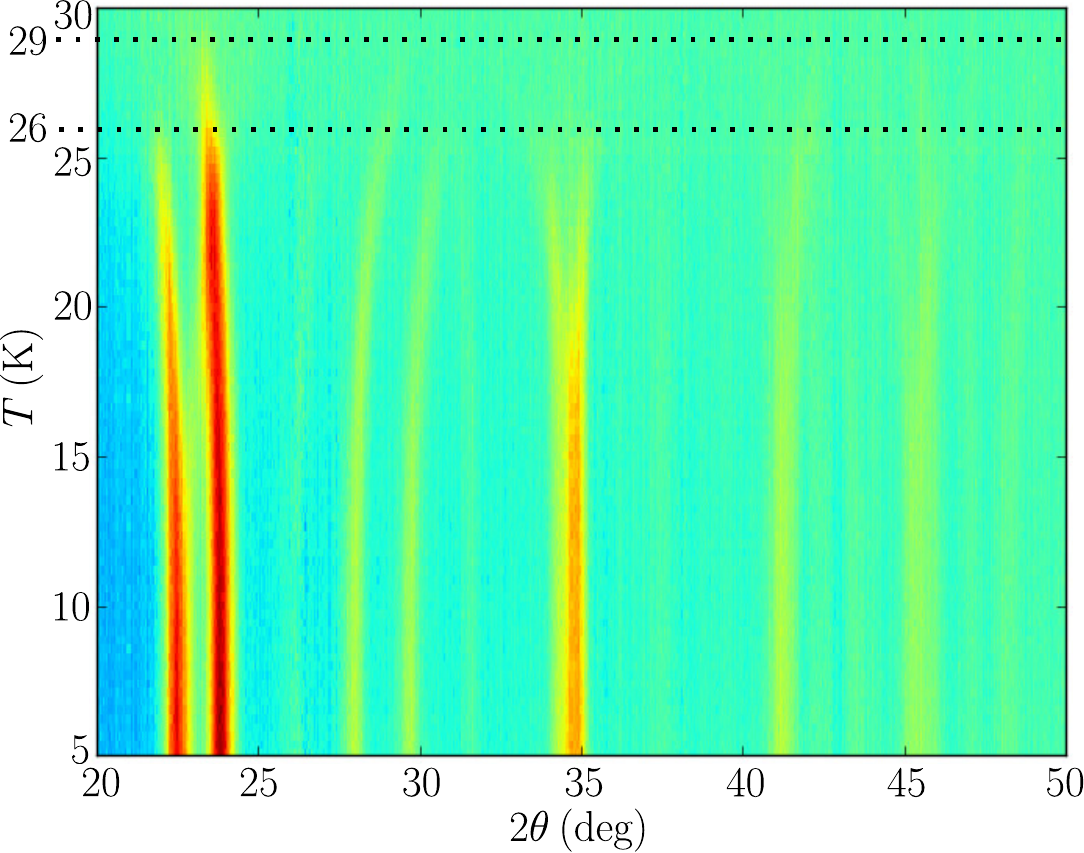}
\caption{\label{thermodiff} NPD data collected at the high-intensity instrument D20 while ramping the temperature. Only the magnetic contribution is shown. The paramagnetic data at 40\,K were used to subtract the nuclear contribution.}
\end{figure}

\begin{figure*}
\includegraphics[angle=0,width=\textwidth]{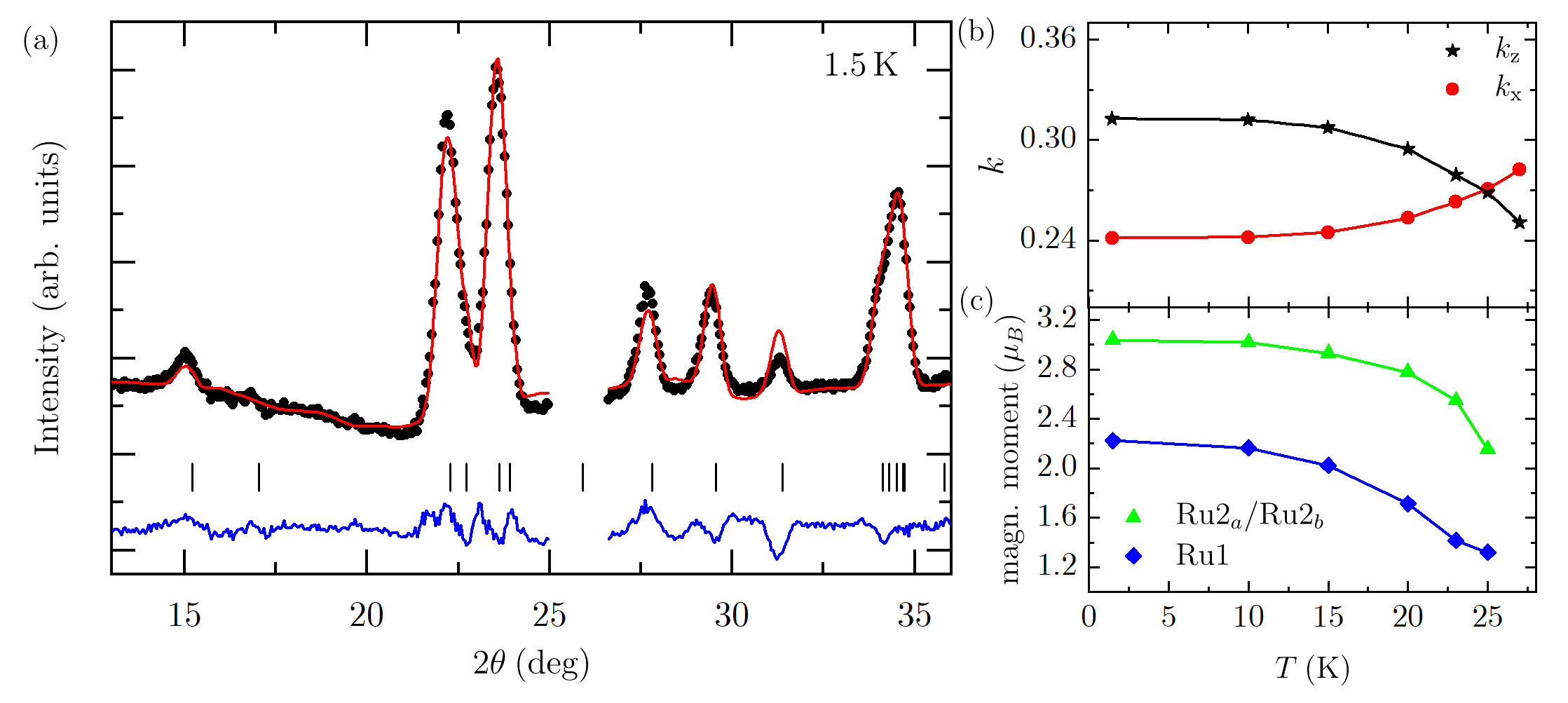}
\caption{\label{D20} Data collected at the high-intensity instrument D20. (a) Magnetic refinement with the \mbox{Rietveld} method for the data (black dots) collected at 1.5\,K with the nuclear background subtracted. The \mbox{Rietveld} fit is depicted in red, the difference between experiment and fit is depicted in blue, and the magnetic peak positions are marked with the black ticks. [(b) and (c)] Evolution of the components of $\vec{k}$ and the magnetic moments for Ru1 and Ru2$_a$/Ru2$_b$ with temperature. The error bars are smaller than the symbol size. The lines are guides for the eye.}
\end{figure*}
\begin{figure*}
\includegraphics[width=0.69\textwidth]{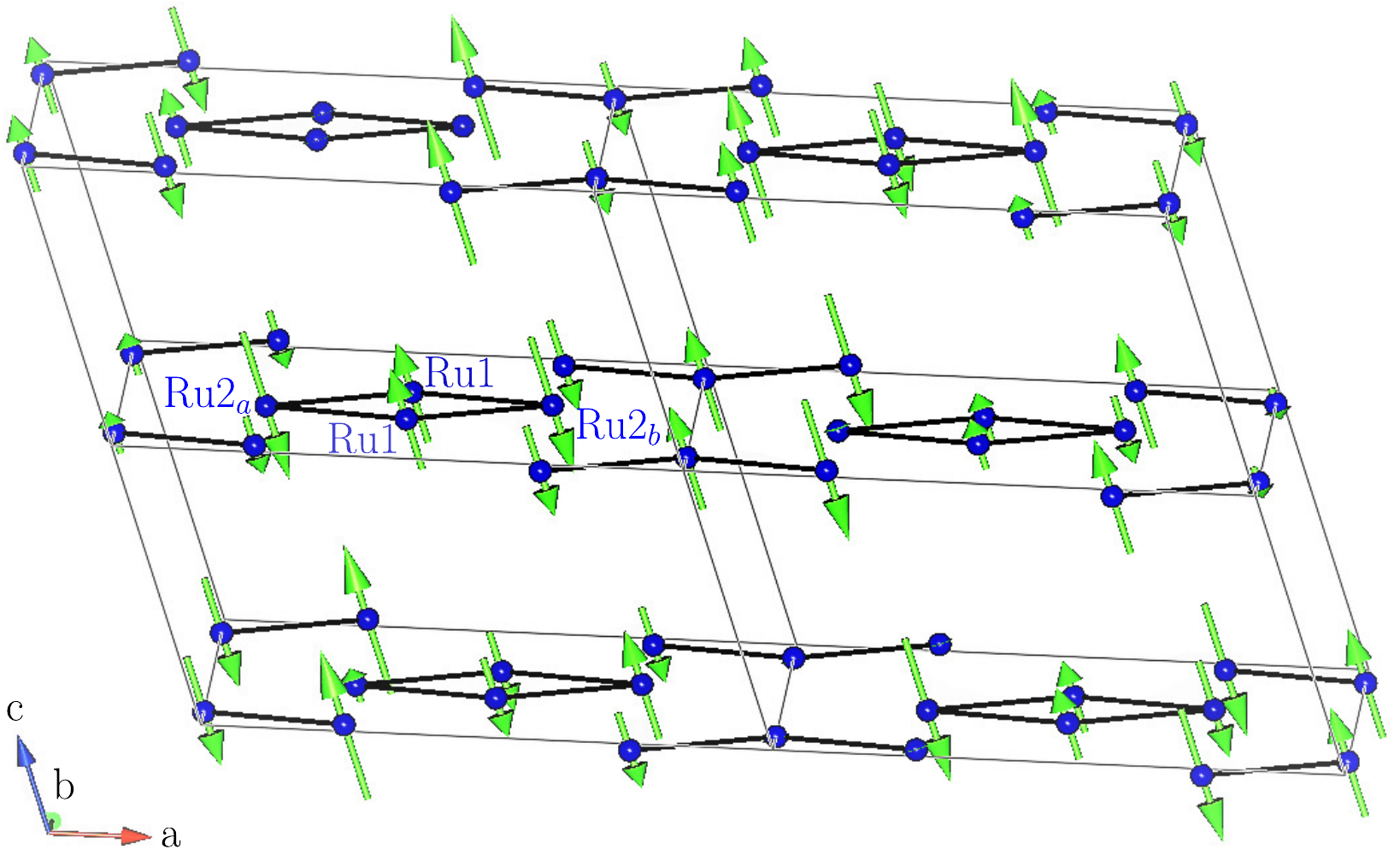}
\caption{\label{magn_structure} Magnetic structure of Na$_3$RuO$_4$ obtained from the magnetic \mbox{Rietveld} refinement. Due to frustrated intertetramer couplings the spin density wave is incommensurately modulated along $a$ and $c$. The magnetic structure was drawn with \texttt{FPStudio}.}
\end{figure*}

To further explore the magnetic structure, we use the data collected at the high-intensity instrument D20. The thermo-diffractogram showing the magnetic contribution after subtracting the nuclear background from the data is pictured in Fig.~\ref{thermodiff}. Upon cooling, magnetic Bragg
peaks start to appear at 29\,K, which fits perfectly to the upper phase transition observed in the heat capacity data. At around 26\,K, a second set of magnetic peaks emerges and parallels the second phase transition in the heat capacity data. Upon further cooling no other peaks appear but the magnetic peaks shift in position.
This already indicates an incommensurately modulated magnetic structure.\\
\indent To capture the magnetic structure quantitatively, measurements at fixed temperatures were recorded and will be analyzed next. Fig.~\ref{D20}(a) shows in black the pattern collected at 1.5\,K after the nuclear background (40\,K data) was subtracted. To determine the propagation vector $\vec{k}=(k_x,k_y,k_z)$, the procedure was as follows. The lattice parameters determined from the the high-resolution D2B data at 40\,K were fixed to refine the wavelength and zero shift for the data from D20. {\cred The possible propagation vectors were determined from the zero-shift corrected positions of the magnetic peaks at 1.5\,K using \texttt{k-SEARCH}. LeBail refinements for the data without the nuclear background were then done using the candidate propagation vectors~\cite{suppl}.} The best solution returned $\vec{k}=(0.242(1),0,0.313(1))$ and the magnetic structure was refined as a spin-density wave with magnetic moments directed along $c$. The corresponding irrep splits the Ru2 site into two, resulting in three nonequivalent Ru sites in the magnetic structure (see Fig.~\ref{magn_structure}). The magnetic \mbox{Rietveld} refinement is shown in Fig.~\ref{D20}(a) as the red line and describes the data accurately with the magnetic R\mbox{-}factor of 4.79.\\
\indent The $x$ and $z$ components of the propagation vector as a function of temperature are depicted in Fig.~\ref{D20}(b). These components are almost temperature-independent at low temperatures and start changing above 15\,K. Whereas $k_x$ increases on heating, $k_z$ decreases. This correlation reflects the same microscopic origin of the incommensurability along $a$ and $c$, as we discuss in more detail below. Between 26 and 29\,K, only a fraction of the magnetic reflections could be observed, and a refinement of the magnetic structure was not possible. However, magnetic peaks remain at incommensurate positions, suggesting that the intermediate ordered phase may be related to the low-temperature one.\\
\indent At low temperatures, the ordered magnetic moments are 2.23(3)\,$\mu_{\text{B}}$ for the Ru atom on site 1 and 3.04(3)\,$\mu_{\text{B}}$ for both Ru atoms on the former site 2. (Without loss of the fit quality, the magnetic moments of the split positions were restricted to be equal.) The determined moments are generally consistent with the theoretically expected local moment of 3\,$\mu_{\text{B}}$ for $S=3/2$. The temperature dependence of the magnetic moments is shown in Fig.~\ref{D20}(c). Below 10\,K, the magnetic moments stay constant within experimental error and then start to decrease monotonically until at 25\,K the moments on Ru1 and Ru2 reach 60\% and 70\% of the 1.5\,K value, respectively.\\
\indent The magnetic structure at 1.5\,K is shown in Fig.~\ref{magn_structure}. The Ru1 and Ru2$_a$/Ru2$_b$ moments are antiparallel within one tetramer. Additionally, a small phase shift of $\phi=0.092(3)$ was found between Ru2$_a$ and Ru2$_b$, but to a first approximation individual tetramers show an almost compensated antiferromagnetic order. The incommensurability arises from the couplings between the tetramers. Below, we show that these intertetramer couplings play the dominant role in the Na$_3$RuO$_4$ magnetism.

\subsubsection{Microscopic Magnetic Model}
Magnetic couplings in Na$_3$RuO$_4$ can be divided into three groups, as shown in Table~\ref{tab:exchange}. The couplings of the first group are within the tetramer. Two of them involve Ru--O--Ru bridges and show a drastically different coupling strength, $J1$ ferromagnetic and $J2$ antiferromagnetic, despite essentially the same Ru--Ru distances. This difference can be traced back to the Ru--O--Ru angles. The coupling $J1$ is closer to the $90^{\circ}$ regime and, therefore, ferromagnetic according to Goodenough-Kanamori-Anderson rules. The coupling $J3$ is weak and involves a Ru--O$\ldots$O--Ru bridge with the O$\ldots$O distance of 2.78\,\r A.

\begin{table}
\setlength{\tabcolsep}{8pt}
\caption{Exchange couplings in Na$_3$RuO$_4$ obtained from \textit{ab initio} calculations: Ru--Ru distances $d$, angles $\varphi$ (Ru--O--Ru and dihedral Ru--O--O--Ru depending on the coupling), and the coupling strength $J$. The labels are introduced in Fig.~\ref{ExCoup}. The couplings not listed in this table are below 1\,K.}
\begin{tabular*}{0.43\textwidth}{ccccc}
    \hline
    bond &label& $d$ (\AA) & $\varphi$ (deg) & $J$ (K) \\
    \hline
    \hline
    \multicolumn{5}{c}{within the tetramer}\\

 Ru1--Ru1 &$J1$ & 3.210 & 98.2 & $-4.9$ \\
 Ru1--Ru2 & $J2$ & 3.210 & 97.5/103.3 & 30.2 \\
 Ru2--Ru2 & $J3$ & 5.559 & 0 & 2.0 \\
 \hline
 \multicolumn{5}{c}{between the tetramers, $ab$ plane}\\
    
 Ru2--Ru2 & $J4$ & 5.477 & 0 & 34.7 \\
 Ru1--Ru2 & $J5$ & 5.531 & 1.6 & 23.4 \\
 Ru1--Ru1 & $J6$ & 6.376 & 28.2 & 7.1 \\
 Ru2--Ru2 & $J7$ & 6.413 & 19.7 & 2.0 \\
\hline
\multicolumn{5}{c}{between the $ab$ planes}\\
    
 Ru1--Ru1 & $J8$ & 5.703 & 74.7 & 6.1 \\
 Ru2--Ru2 & $J9$ & 6.262 & 0 & 20.9 \\
 \hline
  \end{tabular*}
 \label{tab:exchange}
\end{table}
\begin{figure}
\includegraphics[width=0.48\textwidth]{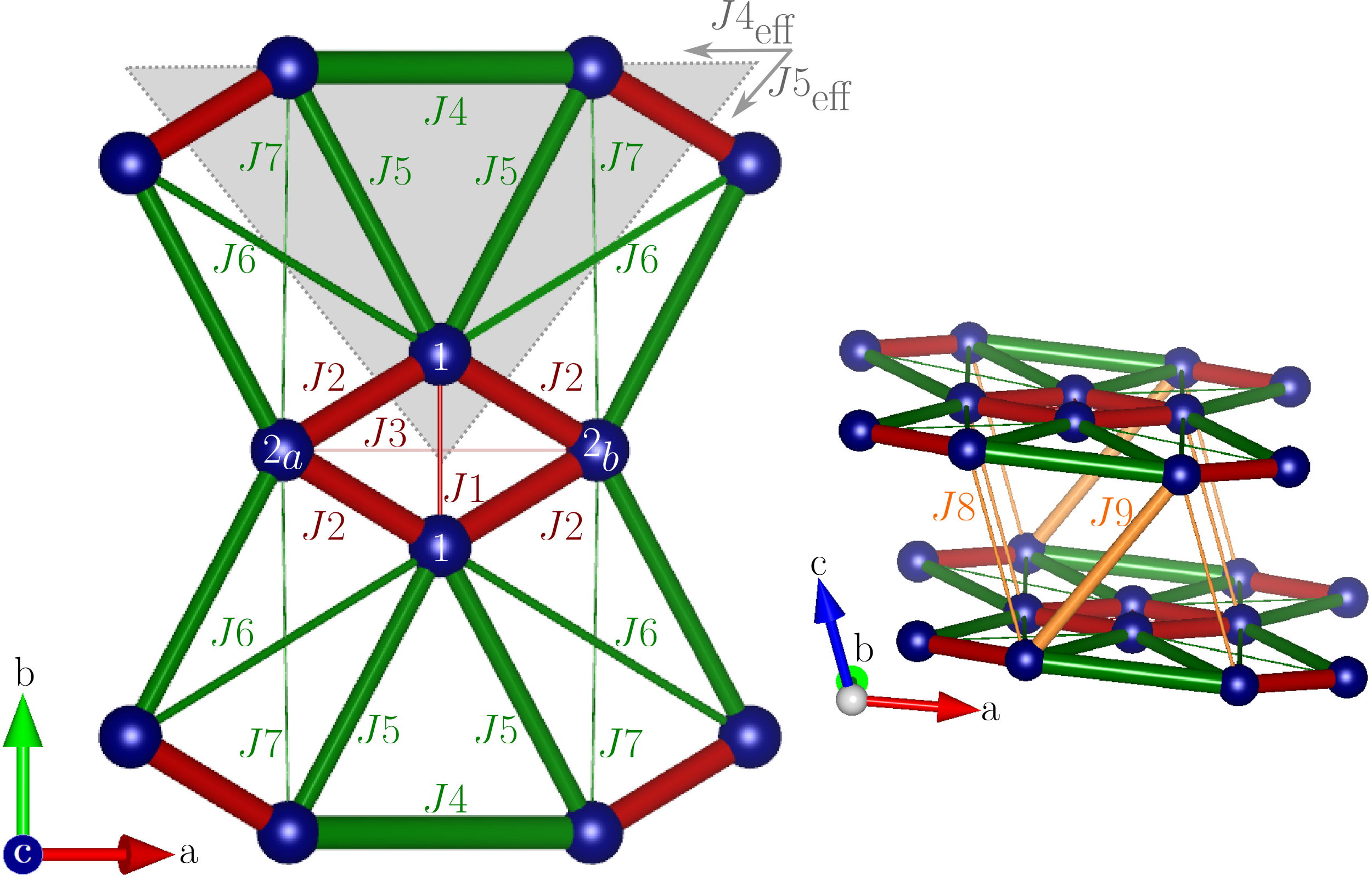}
\caption{\label{ExCoup} Exchange couplings between the Ru atoms. Intratetramer couplings are drawn in red, intertetramer couplings in green, and interlayer couplings in orange. The thickness of the lines shows the strength of the couplings. Exchange couplings not indicated are insignificant. The structures were drawn in \texttt{VESTA} \cite{Vesta2011}.}
\end{figure}
The couplings between the tetramers also involve the Ru--O$\ldots$O--Ru bridges but with the longer O$\ldots$O distances of $3.35-3.65$\,\r A. Some of these couplings are as strong as $J1$ and $J2$. The magnitude of these long-range couplings is controlled by the respective dihedral angles, see Table~\ref{tab:exchange}. The strongest coupling occurs when two Ru atoms and two O atoms all lie in the same plane, similar to long-range superexchange interactions in Cu compounds~\cite{nekrasova2020}. The only exception is $J3$ where the shorter O$\ldots$O distance gives rise to a more curved pathway, which is less suitable for superexchange~\cite{bag2021}. It is worth noting that magnetic interactions in other ruthenium compounds are typically restricted to nearest neighbors with Ru--O--Ru pathways~\cite{suzuki2019}. Na$_3$RuO$_4$ with its strong long-range couplings stands as an exception to this trend.\\
\indent To determine magnetic order stabilized by the couplings $J1-J9$, we performed an energy minimization against three components of the propagation vector ($k_x$, $k_y$, and $k_z$) as well as phase shifts between the magnetic moments within the tetramer ($\alpha_1$, $\alpha_2$, and $\alpha_3$). This {\cred numerical minimization using \texttt{Mathematica}} results in $k_x=0.250$ and $k_z=0.350$ in a good agreement with the experimental $\vec{k}=(0.242(1),0,0.313(1))$ at 1.5\,K. Moreover, we find $\alpha_1=\alpha_2=0.5107$ for the phase shift between Ru1 and Ru2$_a$ as well as $\alpha_3=0.0215$ for the phase shift between Ru2$_a$ and Ru2$_b$. These values also compare favorably with 0.5 and 0.092(3) found experimentally. The good match between the experimental and calculated magnetic structures serves as a verification of our microscopic magnetic model. \\
\indent For a further comparison we calculated effective Curie-Weiss temperatures on both Ru sites as
$$\theta=-\frac{S(S+1)}{3}\sum_i Ji$$
where $S=\frac32$ and the summation is over all couplings that the given Ru site is involved in. This procedure returns the values of $\theta=-161$\,K for Ru1 and $-246$\,K for Ru2 that compare favorably to the experimental Curie-Weiss temperature of $-161.5(5)$\,K. Moreover, the weaker exchange field on Ru1 explains the lower ordered moment on this site [Fig.~\ref{D20}(c)].

\section{Discussion and Summary}
Our microscopic magnetic model shows that individual tetramers in Na$_3$RuO$_4$ are only weakly frustrated. Both ferromagnetic $J1$ and antiferromagnetic $J2$ favor the antiferromagnetic order with opposite magnetic moments on Ru1 and Ru2, whereas $J3$ is too weak to modify this order significantly.\\
\indent The incommensurate order is a consequence of frustrated couplings between the tetramers. Indeed, the two leading intertetramer interactions $J4$ and $J5$ are similar in magnitude and compete. A smaller value of $J4$ would prevent the formation of an incommensurate order. If all the exchange couplings are kept constant and $J4$ is reduced to 26.2\,K or lower, the propagation vector becomes $\vec{k}=(0,0,1/2)$, which corresponds to the picture of antiferromagnetic tetramers that tile up the $ab$ plane. The same applies if $J5$ is raised above 31.9\,K. It turns out that $J4$ and $J5$ in Na$_3$RuO$_4$ are fine-tuned to stabilize incommensurate order. A useful perspective on this frustration mechanism is given by an effective model where each tetramer is reduced to a single site of an anisotropic triangular lattice, as shown by the black dotted lines in Fig.~\ref{structure}(b). For this picture we introduce the effective couplings $J4_{\textrm{eff}}=J4$ and $J5_{\text{eff}}=-2\cdot J5+J6+J7$ for the interactions along $\vec{a}$ and $\vec{a}\pm \vec{b}$, respectively (see Fig.~\ref{ExCoup}). The minus sign in front of $J5$ is due to the fact that this coupling connects Ru1 and Ru2 atoms with antiparallel spins. Since $k_y=0$ in our case, the expression $\cos(\pi \cdot k_x)=-J5_{\text{eff}}/2J4_{\text{eff}}$ ensues. The experimental value of $k_x=0.242$ can be reproduced with $J5_{\textrm{eff}}/J4_{\textrm{eff}}=-1.45$, suggesting a moderate spatial anisotropy of this effective triangular lattice.\\
\indent After elucidating the magnetic order in the $ab$ plane we will now consider the origin of incommensurability along the $c$ direction. Geometric frustration between $J4$, $J8$, and $J9$ alone would not cause this effect because $J8$ is too weak. This can be seen by ignoring other sources of geometric frustration and leaving $J4$ as the only intertetramer interaction in the $ab$ plane. Doing so would result in a commensurate $\vec{k}=(1/2, 0, 0)$ with opposite spin arrangements on the tetramers along $\vec{a}$. Then the dominant interlayer interaction $J9$ creates an opposite spin arrangement of neighboring tetramers along $\vec{a}+\vec{c}$. Given $J8\ll J9$, one can set $J8=0$ and write the $k_z$ contribution to the total energy as $E_{J8=0}\left(k_{\text{z}}\right)=J9\cos\left(2\pi \left(k_{\text{x}}+k_{\text{z}}+\phi \right) \right)$. Because $J9$ is oblique to the $c$ axis, $k_{\text{x}}$ and $k_{\text{z}}$ get linked and the magnetic structure becomes incommensurately modulated along $c$ as a result of the incommensurability in the $ab$ plane. This link is further supported by the correlation between $k_x$ and $k_z$ as a function of temperature [Fig.~\ref{D20}(b)]. The almost symmetric changes in $k_x$ and $k_z$ further support the fact that the incommensurability of $k_z$ is merely a consequence of the oblique coupling $J9$.\\
\indent {\cred Our microscopic model fully explains the magnetic structure of Na$_3$RuO$_4$, including its periodicity that gives rise to the minima of the spin-wave dispersion around $Q=1.0$\,\r A$^{-1}$~\cite{Haraldsen2009}, the position of the two most intense magnetic Bragg peaks in Fig.~\ref{D20}. This maximum of the spin-wave dispersion measured by inelastic neutron scattering~\cite{Haraldsen2009} lies around 6\,meV that corresponds to $J\simeq 2$\,meV in a spin-$\frac32$ triangular antiferromagnet~\cite{Mourigal2013}. This effective $J$-value is indeed comparable to the leading exchange couplings $J4$ and $J5$ (Table~\ref{tab:exchange}). However, further inelastic experiments are needed to reveal details of the spin-wave dispersion and refine magnetic interaction parameters in Na$_3$RuO$_4$ experimentally.} \\
\indent Finally, we comment on the nature of two consecutive phase transitions. We have shown that both transitions are magnetic in nature without any detectable structural component. Our data collected between 26 and 29\,K are unfortunately insufficient to resolve the magnetic structure in this temperature range. However, Fig.~\ref{thermodiff} clearly shows that many of the magnetic peaks persist above 26\,K up to 29\,K, and the magnetic structure remains incommensurate. In this context, an effective description of Na$_3$RuO$_4$ in terms of the anisotropic triangular lattice may be useful because triangular antiferromagnets often show two magnetic transitions~\cite{lee2014a,lee2014b,ranjith2016} with magnetic ordering at $T_{\text{N}}$ followed by spin reorientation. Interestingly, neither of the transitions showed any dependence on the magnetic field up to at least 7\,T [Fig.~\ref{HCMagn}(a)]. This may be due to the quite high energy scale of magnetic interactions and the large local moment ($S=\frac32$). \\
\indent In summary, we have determined the magnetic ground state of Na$_3$RuO$_4$ using NPD. Magnetic structure of this material is incommensurate with the propagation vector $\vec{k}=(0.242(1),0,0.313(1))$ at 1.5\,K. We showed that structural tetramers play only a minor role in the physics because long-range couplings between the tetramers are as strong as the nearest-neighbor coupling within the tetramers. Frustration of the intertetramer couplings renders magnetic structure incommensurate in the $ab$ plane and, consequently, also along the $c$ direction because of the oblique nature of the leading coupling between the planes. This peculiar magnetic model has several ramifications. It puts Na$_3$RuO$_4$ into the category of triangular antiferromagnets rather than molecular (tetramer) magnets, contrary to the previous studies. Moreover, it highlights the importance of long-range Ru--O$\ldots$O--Ru couplings that are typically disregarded in ruthenates. The role of these couplings in other Ru-based magnets, including the honeycomb and pyrochlore systems, remains an interesting avenue for future studies.


Experimental and computational data associated with this study can be found in Refs.~\cite{neutron-data,data}.

\acknowledgments
We acknowledge ILL for providing beamtime for this project. This work was funded by the Deutsche Forschungsgemeinschaft (DFG, German Research Foundation) -- TRR 360 -- 492547816. Computations for this work were done (in part) using resources of the Leipzig University Computing Center.



%

\clearpage\newpage
\begin{widetext}
\begin{center}
\large\textbf{\textit{Supplemental Material}\smallskip \\ Geometrical frustration and incommensurate magnetic order \\ in Na$_3$RuO$_4$ with two triangular motifs}
\end{center}

\renewcommand{\thesection}{S\arabic{section}}
\renewcommand{\thefigure}{S\arabic{figure}}
\renewcommand{\thetable}{S\arabic{table}}
\setcounter{figure}{0}
\setcounter{table}{0}
\setcounter{section}{0}

\section{Crystal structure}

In Tables~\ref{tab:structure}--\ref{tab:structure3}, we list the atomic positions refined from the D2B data. Fig.~\ref{fig:refinement} shows an example of the structure refinement for the 40\,K data. 

Note that the table of the atomic positions in Ref.~\cite{Regan2005} contains two typos. The $z$-coordinate of Na3 should be around 0.52 (general position $8j$), whereas the $y$-coordinate of O3 should be 0.5 when $x\simeq 0.15$ (special position $4i$). This was also pointed out in Ref.~\cite{Haraldsen2009}.

\begin{table}[!h]
\begin{minipage}{13cm}
\caption{\label{tab:structure}
Atomic positions and atomic displacement parameters ($U_{\rm iso}$, in $10^{-2}$\,\r A$^2$) obtained from the D2B data collected at 300\,K. The lattice parameters are $a=11.0286(1)$\,\r A, $b=12.8148(2)$\,\r A, $c=5.7048(1)$\,\r A, and $\beta=109.905(1)^{\circ}$. The refinement residuals are $R_I=0.013$ and $R_p=0.028$. 
}
\begin{ruledtabular}
\begin{tabular}{cc@{\hspace{1.2cm}}cccc}
     &      & $x/a$     & $y/b$     & $z/c$      & $U_{\rm iso}$ \\
 Na1 & $4g$ & 0         & 0.3763(5) &  0         & 0.6(2) \\
 Na2 & $4e$ & 0.25      & 0.25      &  0         & 0.8(2) \\
 Na3 & $8j$ & 0.2464(5) & 0.1238(4) & 0.5218(10) & 0.9(1) \\
 Na4 & $2c$ & 0         & 0         & 0.5        & 0.2(2) \\
 Na5 & $4h$ & 0         & 0.2564(6) & 0.5        & 1.5(2) \\
 Na6 & $2d$ & 0         & 0.5       & 0.5        & 2.0(3) \\
 Ru1 & $4g$ & 0         & 0.1272(2) & 0          & 0.1(1) \\
 Ru2 & $4i$ & 0.2537(3) & 0         & 0.0266(7)  & 0.2(1) \\
 O1  & $4i$ & 0.1094(5) & 0         & 0.1980(9)  & 0.6(1) \\
 O2  & $8j$ & 0.1016(3) & 0.2258(2) & 0.2245(6)  & 0.3(1) \\
 O3  & $4i$ & 0.6570(4) & 0         & 0.2013(8)  & 0.4(1) \\
 O4  & $8j$ & 0.3786(3) & 0.3884(2) & 0.1872(6)  & 0.1(1) \\
 O5  & $8j$ & 0.3530(3) & 0.1071(2) & 0.2320(6)  & 0.7(1) \\
\end{tabular}
\end{ruledtabular}
\end{minipage}
\end{table}

\begin{table}[!h]
\begin{minipage}{13cm}
\caption{\label{tab:structure2}
Atomic positions and atomic displacement parameters ($U_{\rm iso}$, in $10^{-2}$\,\r A$^2$) obtained from the D2B data collected at 40\,K. The lattice parameters are $a=10.9960(2)$\,\r A, $b=12.7769(2)$\,\r A, $c=5.6828(1)$\,\r A, and $\beta=109.832(1)^{\circ}$. The refinement residuals are $R_I=0.012$ and $R_p=0.026$. 
}
\begin{ruledtabular}
\begin{tabular}{cc@{\hspace{1.2cm}}cccc}
     &      & $x/a$     & $y/b$     & $z/c$      & $U_{\rm iso}$ \\
 Na1 & $4g$ & 0         & 0.3757(4) &  0         & 0.1(2) \\
 Na2 & $4e$ & 0.25      & 0.25      &  0         & 0.0(2) \\
 Na3 & $8j$ & 0.2449(4) & 0.1248(3) & 0.5186(8)  & 0.0(1) \\
 Na4 & $2c$ & 0         & 0         & 0.5        & 0.0(2) \\
 Na5 & $4h$ & 0         & 0.2572(4) & 0.5        & 0.7(1) \\
 Na6 & $2d$ & 0         & 0.5       & 0.5        & 0.6(2) \\
 Ru1 & $4g$ & 0         & 0.1267(2) & 0          & 0.0(1) \\
 Ru2 & $4i$ & 0.2545(3) & 0         & 0.0253(6)  & 0.0(1) \\
 O1  & $4i$ & 0.1103(4) & 0         & 0.1979(8)  & 0.3(1) \\
 O2  & $8j$ & 0.1016(2) & 0.2265(2) & 0.2228(5)  & 0.1(1) \\
 O3  & $4i$ & 0.6560(3) & 0         & 0.2012(7)  & 0.0(1) \\
 O4  & $8j$ & 0.3785(2) & 0.3884(2) & 0.1890(5)  & 0.0(2) \\
 O5  & $8j$ & 0.3537(2) & 0.1077(2) & 0.2322(5)  & 0.1(1) \\
\end{tabular}
\end{ruledtabular}
\end{minipage}
\end{table}

\begin{table}
\begin{minipage}{13cm}
\caption{\label{tab:structure3}
Atomic positions and atomic displacement parameters ($U_{\rm iso}$, in $10^{-2}$\,\r A$^2$) obtained from the D2B data collected at 1.5\,K. The lattice parameters are $a=10.9999(2)$\,\r A, $b=12.7749(2)$\,\r A, $c=5.6827(1)$\,\r A, and $\beta=109.826(1)^{\circ}$. The refinement residuals are $R_I=0.012$ and $R_p=0.026$. 
}
\begin{ruledtabular}
\begin{tabular}{cc@{\hspace{1.2cm}}cccc}
     &      & $x/a$     & $y/b$     & $z/c$      & $U_{\rm iso}$ \\
 Na1 & $4g$ & 0         & 0.3760(5) &  0         & 0.2(1) \\
 Na2 & $4e$ & 0.25      & 0.25      &  0         & 0.0(2) \\
 Na3 & $8j$ & 0.2452(4) & 0.1245(3) & 0.5186(8)  & 0.1(1) \\
 Na4 & $2c$ & 0         & 0         & 0.5        & 0.0(2) \\
 Na5 & $4h$ & 0         & 0.2580(4) & 0.5        & 0.5(1) \\
 Na6 & $2d$ & 0         & 0.5       & 0.5        & 0.6(2) \\
 Ru1 & $4g$ & 0         & 0.1269(2) & 0          & 0.0(1) \\
 Ru2 & $4i$ & 0.2541(3) & 0         & 0.0250(6)  & 0.0(1) \\
 O1  & $4i$ & 0.1106(4) & 0         & 0.1966(8)  & 0.3(1) \\
 O2  & $8j$ & 0.1017(2) & 0.2265(2) & 0.2239(5)  & 0.1(1) \\
 O3  & $4i$ & 0.6567(3) & 0         & 0.2020(7)  & 0.0(1) \\
 O4  & $8j$ & 0.3784(2) & 0.3886(2) & 0.1886(5)  & 0.0(2) \\
 O5  & $8j$ & 0.3538(2) & 0.1077(2) & 0.2312(5)  & 0.1(1) \\
\end{tabular}
\end{ruledtabular}
\end{minipage}
\end{table}

\begin{figure}[!h]
\includegraphics{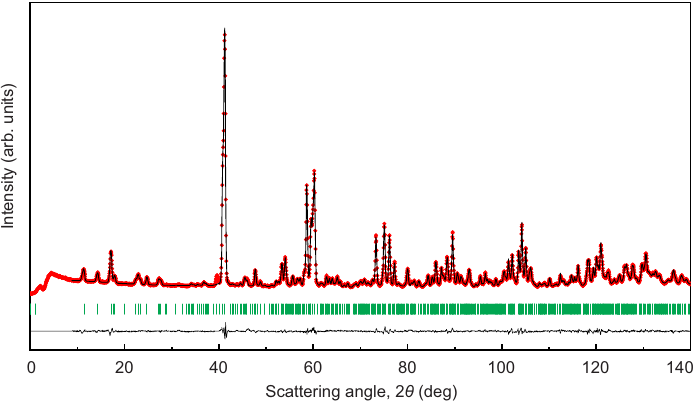}
\caption{\label{fig:refinement}
Rietveld refinement for the D2B data at 40\,K. The small peak from the cryostat window at $2\theta=39^{\circ}$ has been excluded.
}
\end{figure}

\clearpage


\section{Propagation vector}

Running \texttt{k-SEARCH} with 17 detected magnetic Bragg peaks returns three solutions:
\begin{itemize}
 \item $\vec{k}_1=(0.29,0.79,0.19)$
 \item $\vec{k}_2=(0.24,0.01,0.31)$
 \item $\vec{k}_3=(0.99,0.20,0.14)$
\end{itemize}
Le Bail fits of the difference data ($1.5-40$\,K) clearly favor $\vec{k}_2$ as the propagation vector:
\begin{itemize}
 \item $\vec{k}_1:$ $\chi^2=48.3$ and $R_p=8.9$\,\%
 \item $\vec{k}_2:$ $\chi^2=6.4$ and $R_p=3.8$\,\%
 \item $\vec{k}_3:$ $\chi^2=71.7$ and $R_p=11.0$\,\%
\end{itemize}
The corresponding Le Bail fit is shown in Fig.~\ref{fig:lebail}.

\begin{figure}[!h]
\includegraphics{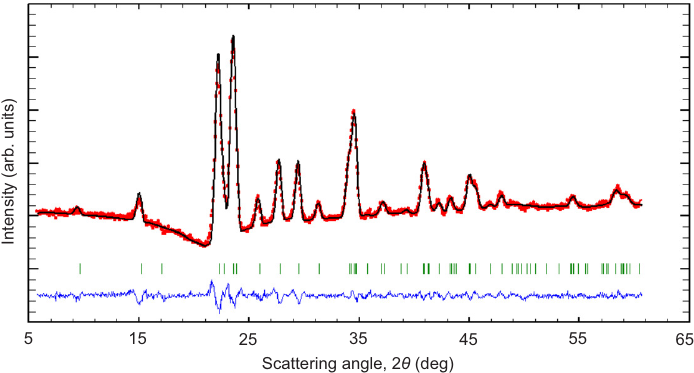}
\caption{\label{fig:lebail}
Le Bail fit of the difference data ($1.5-40$\,K) with the refined propagation vector $\vec{k}_2=(0.242(1),0,0.313(1))$.
}
\end{figure}

\clearpage


\section{\textit{Ab initio} calculations}

Exchange couplings were evaluated in several unit cells:
\begin{itemize}
 \item Cell 1 is the crystallographic unit cell, $\av\times\bv\times\cv$. Here, the couplings $J1$, $J2$, $J3+J4$, $J5$, $J6$, $J7$ can be determined.
 \item Cell 2 is the primitive cell, $(\av+\bv)/2 \times (\av-\bv)/2 \times \cv$. It is twice smaller than cell 1 and does not give access to any additional couplings. It is used for a cross-check of the results.
 \item Cell 3 is the primitive cell doubled along the $c$-direction, $(\av+\bv)/2 \times (\av-\bv)/2 \times 2\cv$, and gives access to the interlayer couplings $J8$ and $J9$.
 \item Cell 4 is the primitive cell expanded in the $ab$-plane, $(\av+\bv) \times (\av-\bv) \times \cv$. It is used to resolve $J3$ and $J4$.
\end{itemize}

\end{widetext}

\end{document}